\documentclass[aps,prd,twocolumn,nofootinbib,showpacs,floatfix]{revtex4}

\usepackage{amssymb,amsfonts}

\usepackage{graphicx}
\usepackage{psfrag}

\newcommand{\be}{\begin{equation}}
\newcommand{\ee}{\end{equation}}
\newcommand{\bea}{\begin{eqnarray}}
\newcommand{\eea}{\end{eqnarray}}
\newcommand{\nn}{\nonumber}

\newcommand{\Hor}{{\mathcal H}}

\newcommand{\DS}{{}^{2}\!D}
\newcommand{\tDS}{{}^{2}\!\tilde D}

\newcommand{\tD}{\tilde D}

\newcommand{\beq}{\begin{equation}}
\newcommand{\eeq}{\end{equation}}

\newfont{\ssg}{cmssbx10 scaled \magstep1}
\newfont{\ssgf}{cmssbx10}

\begin{document}

\title{Numerical implementation of isolated horizon boundary conditions}

\newcommand*{\AEI}{Max-Planck-Institut f\"ur Gravitationsphysik,
  Albert-Einstein-Institut, Am M\"uhlenberg 1, D-14476 Golm, Germany}
\newcommand*{\MEU}{Laboratoire de l'Univers et de ses Th\'eories, UMR 8102
  du C.N.R.S., Observatoire de Paris, F-92195 Meudon Cedex, France}
\newcommand*{\IAA}{Instituto de Astrof\'{\i}sica de Andaluc\'{\i}a, CSIC, Apartado Postal 3004, Granada 
        18080, Spain} 
\newcommand*{\COR}{Center for Radiophysics and Space 
Research, Cornell University, Ithaca, New York, 14853, USA}

\author{Jos\'e Luis Jaramillo}\email{jarama@iaa.es}\affiliation{\MEU\\\IAA}
\author{Marcus Ansorg}\email{marcus.ansorg@aei.mpg.de}\affiliation{\AEI}  
\author{Fran\c cois Limousin}\email{limousin@astro.cornell.edu}\affiliation{\MEU\\\COR}

\date{6 October 2006}

\begin{abstract}
We study the numerical implementation of a set of boundary
conditions derived from the isolated horizon formalism, and 
which characterize a black hole whose horizon is in quasi-equilibrium.
More precisely, we enforce these geometrical
prescriptions as inner boundary conditions on an excised sphere,
in the numerical resolution of the Conformal Thin
Sandwich equations.
As main results, we firstly establish the consistency of including in
the set of boundary conditions a
{\it constant surface gravity} prescription, interpretable as
a lapse boundary condition, and secondly we
assess how the prescriptions presented
recently by Dain et al. for guaranteeing the well-posedness of
the Conformal Transverse Traceless equations 
with quasi-equilibrium horizon conditions 
extend to
the Conformal Thin Sandwich elliptic system.
As a consequence of the latter analysis,
we discuss the freedom of prescribing the expansion
associated with the ingoing null normal at the horizon.

\end{abstract}


\maketitle

\section{Introduction}
The general study of spacetimes containing a black hole whose horizon is
in quasi-equilibrium is of direct interest in astrophysics and
numerical relativity. 
A particularly important application is the determination of inner
boundary conditions 
for the construction of astrophysically realistic 
excised initial data for binary black holes in quasi-circular orbits 
\cite{GourgGB02,Cook02,Pfeif03,CookP04,Ansor05,CaudiCGP06}.
Beyond the construction of initial data, 
the analysis of these boundary conditions also provides a helpful
insight into the evolution problem. More precisely, a generalization
of these conditions constitute an integrant part of the modelization of 
black hole horizons as world-tubes of 
marginally-trapped surfaces, according to the characterizations   
in the quasi-local horizon formalisms of {\it trapped} 
and {\it dynamical horizons}  (see reviews \cite {AshteK05,Booth05}).
In this sense, the quasi-equilibrium conditions here discussed
offer a test-ground for the general dynamical case, in a better
controlled scenario. Lessons acquired in the quasi-equilibrium case
can then be applied to 
the evolution of an excised black hole in a constrained scheme like the one
proposed in Ref. {}\cite{BonazGGN04}.
In addition, quasi-equilibrium conditions themselves are
of direct interest in slow evolution schemes, like the {\it minimal
no-radiation} approximation proposed in Ref. \cite{SchafG04}.

The isolated horizon formalism developed by 
Ashtekar et al. (see Ref. \cite{AshteK05} for a review) provides 
a particularly well-suited framework for studying a black hole
in equilibrium inside a generically dynamical
spacetime\footnote{Throughout the paper we abuse the language and use 
the expression {\it black hole spacetime} to refer to a
spacetime containing an isolated horizon inside, without any mention
to the notion of event horizon.}. It is an example
of trapped horizon in which the horizon world-tube is a null
hypersurface. This null-like character encodes the quasi-equilibrium characterization.

First derivations of quasi-equilibrium horizon boundary conditions were
presented in Refs. \cite{Cook02,CookP04} (see also Ref. \cite{Pfeif03}).
The detailed analysis of isolated horizons in a 3+1 description
of spacetime permits the systematization and extension of these results. 
Following this line of research, an ensemble of boundary conditions for the 3+1 fields 
has been proposed in Refs. \cite{JaramGM04,DainJK05,GourgJ06}. 
In this paper we address the problem of testing numerically these sets
of boundary conditions. 
Translating Einstein equations into a specific system of partial
differential equations, for which the isolated horizon
prescriptions become actual
analytical boundary conditions, requires the choice of a particular
resolution scheme for the geometrical field equations.
For concreteness (and also motivated by the fully-constrained scheme
in Ref. \cite{BonazGGN04}, in which constraints are solved
at each time step), we focus here on the
construction of initial data in a Conformal Thin Sandwich (CTS)
approach \cite{York99,PfeifY03}.
Making use of an {\it excision} technique, i.e. removing a sphere
${\mathcal S}$ of coordinate radius $r_H$ from the initial spatial slice $\Sigma$, and
imposing ${\mathcal S}$ to stand as a space-like slice of an isolated
horizon, an ensemble
of inner boundary conditions for the CTS elliptic system
is determined. Note that throughout the paper we use dimensionless
physical and geometrical quantities which result from the
corresponding dimensional quantities by rescaling with the appropriate
power of $r_H$.  

Regarding index notation, greek letters denote 
lorentzian indices and are mainly used in section
\ref{s:BCgeom}, where null and space-like geometries appear 
in the same context. Latin indices refer specifically to
objects living on space-like slices and are used from section
\ref{s:CTSeqs} on.

\section{Boundary conditions: geometrical form}
\label{s:BCgeom}
\subsection{Geometrical boundary conditions}
In order to introduce a characterization of quasi-equilibrium,
we need a notion of time evolution. A natural evolution vector on a null
world-tube ${\cal H}$, sliced by a given family of trapped surfaces $\{{\cal S}_t\}$, 
is provided by the null-vector $\ell$ that
Lie-draggs ${\cal S}_t$ onto ${\cal S}_{t+\delta t}$. We consider the
horizon slicing $\{{\cal S}_t\}$ as induced by a 3+1 spacetime foliation of 
space-like surfaces $\Sigma_t$. Firstly, we fix the notation. 
We denote by $n^\alpha$ the
time-like unit normal to $\Sigma_t$, by $N$ the associated lapse, 
and by $s^\alpha$ the unit space-like normal to ${\cal S}_t$ lying on 
$\Sigma_t$. The horizon evolution vector
is then expressed as $\ell^\alpha = N \cdot (n^\alpha + s^\alpha)$. The ingoing null
vector $k^\alpha$ (in the plane defined by $n^\alpha$ and $s^\alpha$ and
normalized as $k^\mu \ell_\mu=-1$) is written 
as  $k^\alpha = \frac{1}{2N} (n^\alpha - s^\alpha)$. We denote by
$(\gamma_{\alpha\beta},
K^{\alpha\beta})$ the 3-metric on $\Sigma_t$ and the extrinsic curvature
(with sign convention $K_{\alpha\beta}=-\frac{1}{2}{\cal
L}_n\gamma_{\alpha\beta}= -{\gamma^\mu}_\alpha \nabla_\mu
n_\beta$). The induced metric on the marginally trapped surface ${\cal S}$ is
then given by $q_{\alpha\beta} = \gamma_{\alpha\beta} - s_\alpha
s_\beta$. \\
\noindent Quasi-equilibrium boundary conditions follow from
prescribing certain 3+1 fields to be {\it time independent} on
the horizon. In addition, other relevant boundary conditions (not necessarily
related to quasi-equilibrium) follow
from: i) analytical requirements on the well-posedness of the elliptic
system, ii) numerical control of the horizon slicing taking into account the
geometry of the horizon and, iii) choice of coordinate system adapted
to the horizon (ultimately motivated by numerical reasons).
We briefly review this ensemble of conditions (for their systematic
derivations and justifications, see Refs. \cite{JaramGM04,DainJK05,GourgJ06}).

\subsection {Quasi-equilibrium conditions}
Prescribing the
time-independence of a particular combination of
3+1 fields can represent either an actual {\it restriction} on the 
geometry of ${\cal H}$ as a spacetime hypersurface, or rather
it can refer to the manner ${\cal H}$ is {\it described} in the 3+1
slicing. Both cases are relevant in a numerical relativity context.

1. The minimal notion of quasi-equilibrium is provided by 
the {\it non-expanding horizon} condition, namely the time
independence of the induced metric on ${\cal S}$: 
${q^\mu}_\alpha {q^\nu}_\beta{\cal L}_\ell q_{\mu\nu}=0$. 
Expressed in terms of the expansion $\theta_{(\ell)}$ 
and shear $\sigma_{(\ell)}$ associated with the 
outgoing null normal $\ell^\alpha$, i.e.
\bea
\theta_{(\ell)}&=&
q^{\mu\nu}\nabla_{\mu}\ell_\nu \ \ , \\
(\sigma_{(\ell)})_{\alpha\beta} &=& 
{q^\mu}_\alpha {q^\nu}_\beta \nabla_{\mu}\ell_\nu
-\frac{1}{2}\theta_{(\ell)} q_{\alpha\beta} \ \ ,
\eea
this amounts to (\cite{DreyeKSS03,CookP04,JaramGM04}; see also Ref. \cite{Cook02} for a
heuristic discussion on the vanishing of the shear)
\bea
\theta_{(\ell)} = 0 = \left(\sigma_{(\ell)}\right)_{\alpha\beta} \ \ . \label{NEHgeom}
\eea
These three conditions [$\sigma_{(\ell)}$
is a symmetric trace-less tensor on $S^2$] constitute an actual restriction on the geometry of ${\cal H}$,
essentially related to its null-character via the Raychaudhuri
equation. Physically, they mean that the area of the horizon
remains constant in time. 

2. Another proposed quasi-equilibrium condition \cite{JaramGM04} 
consists in prescribing the time-independence of the
vertical (i.e. in the $\ell^\alpha$ direction) component of the angular
variation of the null normal $\ell^\alpha$. 
Explicitly this translates into ${\cal L}_\ell \Omega_{\alpha} =
0$, where $\Omega_{\alpha} := -
{q^\mu}_{\alpha}\left(k_\nu\nabla_\mu \ell^\nu\right)$. 
Recasting $\Omega_{\alpha}$ in terms of 3+1 fields, 
this condition means that the combination $\Omega_\alpha = {}^2D_\alpha \mathrm{ln} N  - K_{\mu\nu} s^\mu
q^\nu_{\ \, \alpha}$ remains constant in time
(where  ${}^2D$ is the connection associated with
$q_{\alpha\beta}$). Its underlying justification uses the 
weakly isolated horizon notion (see Ref. \cite{JaramGM04}). However, we can
heuristically motivate it in two manners: \\
\noindent {\it i)}  Given ${\cal S}_t$ with an axial symmetry
generated by $\phi^\mu$
and with volume element ${}^2\epsilon=\sqrt{q} \,dx^a\!\!\!\wedge \!dx^b$, 
an angular momentum 
$J_{\cal H} = -\frac{1}{8\pi G}\int_{{\cal S}_t} \Omega_\mu \phi^\mu
\; {}^2\epsilon$ can be 
associated with the horizon \cite{AshteBL01}.
The  surface density of this
momentum, $\Omega_\alpha$, satisfies a Navier-Stokes-like evolution equation  
for a viscous fluid (see \cite{GourgJ06,Gourg05} for a discussion of each term):
\bea
{\cal L}_\ell \Omega_\alpha
 + \theta \Omega_\alpha &=&  8\pi q^\mu_{\ \, \alpha} T_{\mu\nu} \,
 \ell^\nu + 
{}^2D_\alpha \kappa \nn \\  
  &-& \DS_\mu \sigma^\mu_{\ \, \alpha}
  + \frac{1}{2} \DS_\alpha \theta \ \ , \label{Navier-Stokes}
\eea
where $\kappa$ is the non-affinity coefficient of the null geodesic
generated by $\ell$: $\nabla_\ell \ell^\alpha = \kappa \;\ell^\alpha$.
Non-expanding conditions in Eq. (\ref{NEHgeom}) imply the vanishing of the viscous 
terms in Eq. (\ref{Navier-Stokes}) (as well as the {\it force surface
  density} term; see \cite{GourgJ06}), which 
reduces then to the Euler equation, ${\cal L}_\ell
\Omega_{\alpha}={}^2D_\alpha \kappa$.
A natural quasi-equilibrium condition in this fluid analogy is given by
${\cal L}_\ell \Omega_{\alpha}= 0$, which translates 
into: $ {}^2D_\alpha \kappa=0$.
This expresses the constancy of $\kappa$
on ${\cal S}$. Writing then $\kappa = \kappa_o=
\mathrm{const}$, its 3+1 decomposition provides an evolution 
equation for the lapse: 
\bea
{\cal L}_\ell \mathrm{ln} N = \kappa_o - s^\mu D_\mu N 
   + N K_{\mu\nu} s^\mu s^\nu \ \ .
\eea
Under the gauge choice  ${\cal L}_\ell N = 0$, this equation becomes 
the boundary condition proposed in Ref. \cite{JaramGM04}:
\bea
\kappa_o = s^\mu D_\mu N  - N K_{\mu\nu} s^\mu s^\nu \label{kappa=const} \ \ .
\eea
This condition does not restrict the geometry (and therefore 
the physical features) of the horizon, but rather chooses a convenient
3+1 description.
 
\noindent {\it ii)} From the relation above 
 ${\cal L}_\ell \Omega_{\alpha}= 0$, one can 
motivate the time-independence of $\Omega_{\alpha}$
from the Hamiltonian analysis of isolated horizons
(see \cite{AshteFK00,AshteK05}). In this context, 
the non-affinity coefficient $\kappa$ is interpreted 
as the {\it surface gravity} of the horizon. The constancy of 
$\kappa$ turns out to be a very natural equilibrium condition as the quasi-local
zeroth law of black hole mechanics. Such a Hamiltonian analysis 
provides a canonical constant value for the surface gravity, 
namely $\kappa_o=\kappa_{_{\mathrm{Kerr}}}(a,J)$, where
$a$ and $J$ are the area and angular momentum of 
${\cal S}_t$, and $\kappa_{_{\mathrm{Kerr}}}(a,J)$ is the 
corresponding surface gravity of a Kerr black hole.

\subsection{Other geometrical boundary conditions}
Together with quasi-equilibrium motivations, some geometrical
inner boundary conditions follow from genuine numerical motivations.
As a first instance of such boundary conditions, in some numerical 
schemes it is important to keep the location 
of the horizon fixed at a given position. Geometrically this means
that the 3+1 evolution vector $t^\alpha=Nn^\alpha+\beta^\alpha$ (where $\beta^\alpha$
is the shift vector) must be tangent to the horizon
hypersurface ${\cal H}$. Decomposing the shift in its normal and 
tangential parts to ${\cal S}$, i.e. $\beta^\alpha = bs^\alpha - V^\alpha$ (with
$V^\mu s_\mu=0$), it follows $t^\alpha = \ell^\alpha - V^\alpha +
(b-N)s^\alpha$. 
Therefore $t^\alpha$ is tangent to ${\cal H}$ iff
\bea
b-N = 0 \label{b=N} \ \ .
\eea
A second example of numerically motivated boundary condition follows 
from the need of a prescription for the lapse 
that incorporates information on the geometrical content of the horizon, 
but still leaves some rescaling freedom to control
the {\it magnitude} of $N$.
In other words, a geometrical prescription for the slicing $\{{\cal S}_t\}$ of 
${\cal H}$, rather than for $N$ itself. 
In Ref. \cite{AshteBL02} it is shown that specifying
${}^2D^\mu\Omega_\mu$, i.e. the divergence
of the 1-form $\Omega_\mu$ introduced above, 
fixes  the slicing of ${\cal H}$ 
in an intrinsic manner (see also Ref. \cite{GourgJ06}
for a 3+1 discussion). From the 3+1 expression of $\Omega_\alpha$ it follows
\bea
{}^2\Delta \mathrm{ln} N = \DS^\rho (q^\mu_{\ \, \rho} K_{\mu\nu} s^\nu) 
+ {}^2D^\mu\Omega_\mu \ \ , \label{Domega}
\eea
which fixes $\mathrm{ln} N $ up to a constant. In consequence,
the lapse is specified modulo a multiplicative constant that
can be chosen to keep the slicing under numerical control.

\section{Conformal Thin Sandwich decomposition}
\label{s:CTSeqs}
\subsection{CTS Equations}
Fixing a representative\footnote{We use latin indices from now on to
 emphasize that equations are defined on a space-like slice $\Sigma$.}
 $\tilde{\gamma}_{ij}$ in the conformal class
of $\gamma_{ij}$, we perform the following decomposition of the 3-metric
and the extrinsic curvature
\bea
\label{e:conformal_decomposition}
\gamma_{ij} = \Psi^4 \tilde{\gamma}_{ij} \ \ , \ \
K^{ij}=\Psi^{-4}\left(\tilde{A}^{ij}+\frac{1}{3}K\gamma^{ij}\right) \ \ ,
\eea
where 
\bea
\tilde{A}^{ij}=  \frac{1}{2N}\left[(\tilde{L}\beta)^{ij} +
\dot{\tilde{\gamma}}^{ij}\right] \ \ , \ \ K=\gamma^{ij}K_{ij},
\eea 
with $\dot{\tilde{\gamma}}^{ij}:={\mathcal L}_t\tilde{\gamma}^{ij}$, and
\bea
(\tilde{L}\beta)^{ij}=\left(\tilde{D}^i\beta^j + \tilde{D}^j\beta^i
-\frac{2}{3} \tilde{D}_k\beta^k \tilde{\gamma}^{ij}\right) 
\eea
and $\tilde{D}_i$ the Levi-Civita connection associated with $\tilde{\gamma}_{ij}$.
Inserting this decomposition in the Hamiltonian and momentum
constraints, and prescribing on $\Sigma$ the time derivative 
of the trace of the extrinsic curvature ($\dot{K}:= {\mathcal L}_t K$), results
in the CTS equations. 
This is an elliptic system for
$(\Psi,\beta^i,N)$, once the initial {\it free data}  
$(\tilde{\gamma}_{ij},K,\dot{\tilde{\gamma}}^{ij},\dot{K})$ are prescribed 
on the initial $\Sigma$. If a sphere ${\cal S}$ is excised in
$\Sigma$, as it is the case here, 
the boundary conditions for $(\Psi,\beta^i,N)$ on ${\cal S}$ are also
a part of the free data. For concreteness, in this work we focus on 
free data\footnote{\label{b=Nquasi-eq} This choice is used in the
  literature \cite{GourgGB02,Cook02,CookP04,Ansor05,CaudiCGP06} as a
quasi-equilibrium prescription in the {\it bulk}. Gauge 
horizon condition (\ref{b=N}) gains then geometrical meaning, since
it links bulk and horizon quasi-equilibrium notions by making 
the evolution vector $t^\alpha$ to be tangent to the horizon.}
with $\dot{\tilde{\gamma}}^{ij}=0=\dot{K}$, and consider different
possibilities for $(\tilde{\gamma}_{ij},K)$. Vacuum CTS equations,
after a conformal rescaling of the lapse $N=\Psi^a\tilde{N}$, have the form
\bea
&&\displaystyle \tilde{\Delta} \Psi \!-\! \frac{\tilde{R}}{8}\Psi \!
+ \! \frac{1}{32}\Psi^{5-2a}\tilde{N}^{-2}(\tilde{L}\beta)_{ij}(\tilde{L}\beta)^{ij}
\!-\! \frac{1}{12} K^2\Psi^5 \!=0,   \nonumber \label{eqPsi}\\
&&\displaystyle \tilde{\Delta} \beta^i+\frac{1}{3} \tD^i\tD_k\beta^k +
{\tilde{R}^i}_k\beta^k -\tilde{N}^{-1} (\tilde{L}\beta)^{ik}\tilde{D}_k\tilde{N}  \nonumber \\
&&\displaystyle \qquad -(a-6) \Psi^{-1} (\tilde{L}\beta)^{ik}\tilde{D}_k\Psi 
= \frac{4}{3}\Psi^a\tilde{N}\tD^iK  \ \ , \label{CTSeqs} \\
&&\displaystyle \tilde{\Delta}\tilde{N}+2(a+1)\tD^k\mathrm{ln}\Psi\tD_k\mathrm{ln}\tilde{N}   \nonumber \\ 
&&\displaystyle+\tilde{N}\left[\frac{a}{8}\tilde{R} + \frac{a-4}{12}\Psi^4K^2 
+a(a+1)\tD^k\mathrm{ln}\Psi\tD_k\mathrm{ln}\Psi\right]  \nonumber \\
&&\displaystyle-\frac{a+8}{32} \Psi^{4-2a}\tilde{N}^{-1}(\tilde{L}\beta)_{ij}(\tilde{L}\beta)^{ij} =
\Psi^{4-a} \beta^k\tD_k K  \nonumber \label{eqN}  \ \ .
\eea
Different choices of the rescaling exponent $a$
have been  considered in the literature: $a=6$ for defining the {\it conformal lapse}
in Refs. \cite{York99,PfeifY03}, 
$a=-1$ is used in the numerical implementations of Refs. \cite{CookP04,Ansor05,CaudiCGP06}, 
and $a=-2$ in Ref. \cite{BonazGGN04}.
The problem we address in this work is the numerical study of the 
elliptic system (\ref{CTSeqs}), when completed 
with the inner boundary conditions in Sec. \ref{s:BCgeom}.

\subsection{Inner boundary conditions}
Inserting the 
conformal decompositions\footnote{We also introduce the rescaled
induced metric $\tilde{q}_{ij}$ on ${\cal S}_t$: 
$\tilde{q}_{ij} = \Psi^4 q_{ij} = 
\tilde{\gamma}_{ij} - \tilde{s}_i
\tilde{s}_j$, with $\tilde{s}^i = \Psi^2 s^i$.} 
(\ref{e:conformal_decomposition}) in the
geometrical boundary conditions (\ref{NEHgeom}), one finds 
\bea
\label{e:BC_Psi}
\displaystyle 4\tilde{s}^i\tD_i\Psi + \tD_i\tilde{s}^i \Psi = &-&\frac{\Psi^{3-a}}{2\tilde{N}}
(\tilde{L}\beta)_{ij}\tilde{s}^i\tilde{s}^j \\
&+& \Psi^3K\left(1 -\frac{1}{3}\tilde{\gamma}_{ij}\tilde{s}^i\tilde{s}^j\right) \nonumber \ \ ,
\eea
from the vanishing expansion \cite{Thorn87,Cook02,JaramGM04,Dain04,Maxwe04}, and 
\bea
\label{e:shearzero}
\displaystyle
&&\left[\tDS^i V^j + \tDS^j V^i
    - (\tDS_k V^k)\,  \tilde q^{ij}\right]  \\
\displaystyle
&+&\left[\left(\tilde{N}\Psi^{a-2} - \tilde b\right)
\left(\tilde{H}^{ij} -\frac{1}{2} \tilde{q}^{ij}
\tilde{H}\right)\right] = 0 \ \ ,\nonumber
\eea
from the vanishing of the shear \cite{CookP04,JaramGM04}, where
$\beta^i = \tilde{b} \tilde{s}^i - V^i$,
$\tilde{H}_{ij} =  {q^k}_i 
\tilde{D}_{k}\tilde{s}_j$ is the (conformal) extrinsic
curvature of ${\mathcal S}$ as a hypersurface in
$\Sigma$ (with trace $\tilde{H}=q^{kl}\tilde{H}_{kl}$),
and we have used $\dot{\tilde{\gamma}}^{ij}=0$. 
In case of enforcing boundary condition (\ref{b=N}), the
second bracket cancels, and the vanishing of the first line
of Eq. (\ref{e:shearzero}) characterizes $V^i$ as a
{\it conformal Killing vector} of $({\mathcal
  S}_t,\tilde{q}_{ij})$ \cite{CookP04,JaramGM04}. This provides a 
Dirichlet condition for $V^i$ once the conformal isometry has been
chosen. 

In addition to the geometrical or numerical motivations
for the boundary conditions, we must also consider at this point
the analytical well-posedness of the elliptic problem. A
procedure to establish the uniqueness of the solution
of an elliptic equation (even though not straightforwardly 
generalizable to an elliptic system)
consists in making use of a  maximum 
principle \cite{Maxwe04,Dain04}.
This involves the control of the convexity of the functions 
we are solving for (in our case depending on the choice of the exponent $a$), and in particular on  
the signs of their radial derivatives at the boundaries. 
From condition (\ref{e:BC_Psi}), control of $\tilde{s}^i\tD_i\Psi$ on the horizon
demands the control on the sign and size of $(\tilde{L}\beta)_{ij}\tilde{s}^i\tilde{s}^j$, 
something  problematic if only the Dirichlet condition (\ref{b=N}) is imposed. An alternative
condition proposed in \cite{DainJK05}, in the context of a Conformal
Transverse Traceless (CTT) decomposition,
consists in prescribing \footnote{Notation here differs from that in
  Ref. \cite{DainJK05}. Tilded objects there represents physical
  quantities, whereas here they refer to conformal counterparts
  according to (\ref{e:conformal_decomposition}).} $\Psi^6\cdot K_{ij}s^is^j$ to satisfy 
\bea
\label{Kss_ineq}
-\tilde{H}<\Psi^6\cdot K_{ij}s^is^j\leq 0 \ \ .
\eea
Denoting $f_1\equiv\Psi^6\cdot K_{ij}s^is^j$ this is enforced as 
\bea
\label{e:BC_btilde_mixed}
2\tilde s^k\tD_k \tilde{b} - \tilde{b} \tilde H &=& 
3 N \Psi^{-6} f_1 - \tDS_k V^k \nn \\
&-& 2 V^k \, \tilde{D}_{\tilde{s}} \tilde{s}_k -N K \ \ . 
\eea
If condition (\ref{b=N}) is not imposed, the vanishing of the shear  must be fulfilled
by choosing, in addition to $V^i$ as a conformal Killing symmetry, 
free data such that the traceless part
of $\tilde{H}_{ij}$ vanishes ({\it umbilical condition}).
More generally one could solve condition (\ref{e:shearzero})
as an equation for $V^i$ \cite{GourgJ06}. 
Extending the well-posedness
analysis from the CTT to the CTS case is
an important issue.
Difficulties are two-fold. On the one hand, as pointed out in \cite{PfeifY05},
signs in Eqs. (\ref{CTSeqs}) 
for the $(\tilde{L}\beta)_{ij}(\tilde{L}\beta)^{ij}$ terms in the $\Psi$ 
and $\tilde{N}$ 
are problematic for applying a
maximum principle argument. 
No obvious choice of $a$ cures the problem.
On the other hand, the inclusion of new non-linear coupled boundary conditions for $\tilde{N}$
makes the analytical problem even harder. However, a strong motivation for 
boundary condition (\ref{Kss_ineq}) follows from its close relation
with  the characterization of  ${\cal S}_t$ as a
{\it future} trapped surface, i.e. with $\theta_{(k)}\leq 0$, via the identity
\bea
\label{e:BC:fut_marg_trapped}
K_{ij}s^is^j-K=\frac{\theta_{(\ell)}}{2N} + N\theta_{(k)} =
N\theta_{(k)}=\theta_{(\hat{k})}   ,
\eea
where $\hat{k}^\alpha = \frac{1}{2}(n^\alpha - s^\alpha)$. 
More importantly, condition (\ref{Kss_ineq}) is not specifically tied to quasi-equilibrium;
in fact, $\theta_{(k)}\leq 0$ is part in the very definition of 
{\it dynamical horizons}, 
in order to garantee the horizon area increase law \cite{AshteK05}.
Moreover, such a condition on the sign of $\theta_{(k)}$ permits 
to exclude certain {\it pathologies} in the evolution of the horizon,
e.g. the appearance of self-intersections \cite{Simon06} of the
surface ${\cal S}_t$, something to be avoided during the non-merger
phase of the black hole evolution. 
The present quasi-equilibrium context offers a controlled test-bed
for studying this condition. 
Numerical techniques seem to be an
appropriate tool for a first analysis of this problem.

\section{Numerical results}
We make use of two independent codes for solving Eqs. (\ref{CTSeqs}),
both using pseudo-spectral methods: the first one employs the elliptic
solvers described in Refs. \cite{BonazGM99,GrandNGM01,GrandGBO2} and
implemented in the C++ library LORENE \cite{Loren}. 
The second code has been specifically designed for the purpose of this
paper. It uses a single domain technique for solving
elliptic boundary value problems in the exterior of an excised
spherical shell.

In order to determine the elliptic
system, we complete Eqs.  (\ref{CTSeqs}) with a
specific combination of five of the boundary conditions reviewed 
in the previous section. 
Different possibilities arise, but all of them must 
incorporate Eq. (\ref{e:BC_Psi}) and Eq. (\ref{e:shearzero}).
For concreteness, in this paper we will restrict ourselves to 
axially symmetric excised spheres (with azimuthal symmetry $\phi^i$), 
and impose as a Dirichlet boundary condition for the tangential part 
of the shift: $V^i = \Omega_o \cdot \ \phi^i$, with $\Omega_o$ a constant
[vanishing of the shear requires then appropriate additional boundary conditions
or suitable free initial data in order to cancel the second line in Eq. (\ref{e:shearzero})].
Eqs. (\ref{e:BC_Psi}) and (\ref{e:shearzero}) 
fix three of the five boundary conditions. The remaining two will be chosen among 
Eqs. (\ref{kappa=const}), (\ref{b=N}), (\ref{Domega}) and (\ref{e:BC_btilde_mixed}).
At this stage of the analysis it is
methodologically useful to interpret each condition as associated with
a specific equation of the system, even if this makes no strict sense due to the 
(non-linear) coupled character of the boundary conditions.
Table \ref{t:method_BC} summarizes this strategy, followed in Refs. \cite{Cook02,CookP04,JaramGM04,DainJK05}.
\begin{table}
\begin{center}
\begin{tabular}{|c|c|c|c|}
\hline 
$\Psi$ & $V^i$ &  $\tilde{b}=\beta^k \tilde{s}_k$& $N$  \\
\hline 
$\theta_{(\ell)}=0$  & $V^i = \Omega_o \cdot \ \phi^i$ & 
$b=N$ & 
  $\kappa =\mathrm{const}$ \\
&& $\Psi^6\cdot K_{ij}s^is^j = f_1$ & ${}^2D_k\Omega^k = f_2$ \\
\hline
\end{tabular}
\end{center}
\caption[]{\label{t:method_BC}
Methodological assignment of boundary conditions to constrained fields.  
We keep fixed conditions for $\Psi$ and $V^i$ and study different
combinations 
of the
conditions for $\tilde{b}$ and $N$.   
}
\end{table}
The numerical implementations show that, keeping fixed the conditions {\it for} $\Psi$ and $V^i$, 
the different combinations
of conditions for  $\tilde{b}$ and $N$ actually lead to the 
existence of solutions, at least for 
appropriate ranges of the free parameters (and independently of 
$\tilde{\gamma}_{ij}$ and $K$). The solutions
are {\it generically} unique or finite in number, and only very particular 
({\it non-generic}) choices of the functions $f_1$ and $f_2$ lead to the appearance of an
infinite number of solutions.

\subsection{ Combination ($b=N, \kappa =\mathrm{const}$)}
This set of boundary conditions implements the vanishing of the shear
for any choice of $\tilde{\gamma}_{ij}$, as well as the constancy of the surface gravity.
In Ref. \cite{JaramGM04} it was proposed the Hamiltonian canonical choice
$\kappa_o=\kappa_{_{\mathrm{Kerr}}}(a,J)$, i.e.
\bea
s^iD_iN -NK_{ij}s^is^j=\kappa_{_{\mathrm{Kerr}}}(a,J) \ \ ,\label{kappaKerr}
\eea
where $\kappa_{_{\mathrm{Kerr}}}(a,J)$ is a functional of $(\Psi, \beta^i, N)$. 
However, the resulting elliptic system presents some problematic features.
We consider firstly the spherically symmetric case, $\Omega_o=0$. 
The elliptic system admits then an infinite number of solutions, 
since maximal slicings of Schwarzschild provide the 
2-parameter Estabrook-Wahlquist family of solutions \cite{EstabW73}, as pointed out in 
Ref. \cite{CookP04}.
Fixing the 
coordinate radius of the excised sphere only fixes one of these parameters, leading
to a degenerated problem. 
The very nature of this degeneracy 
suggests the way out,
since setting the surface gravity to a {\it given} constant $\kappa_o$
fixes the representative of the Estabrook-Wahlquist family. As a code test,
we have confirmed numerically that the 
system ($b=N, \kappa = \kappa_o$) determines a unique solution,
leading to a well-posed problem.
In addition, the {\it a posteriori} evaluation of the quantity
$\kappa_{_{\mathrm{Kerr}}}(a,J)$ on the constructed solution,
results in $\kappa_{_{\mathrm{Kerr}}}(a,J)=\kappa_o$. This is in agreement with the
degeneracy of the system [$b=N, \kappa_o=\kappa_{_{\mathrm{Kerr}}}(a,J)$], 
meaning that the operators on the left- and right-hand sides of Eq. (\ref{kappaKerr}) 
become identical on the space of solutions of this elliptic system. 
We conclude that, in the spherically symmetric case, a well-posed
problem is defined by imposing the
coordinate system to be adapted to the horizon ($b=N$) together with 
prescribing a given constant value for the
surface gravity.
In this system, the Hamiltonian canonical value for the surface gravity 
$\kappa =\kappa_{_{\mathrm{Kerr}}}(a,J)$ cannot be imposed {\it as a boundary condition},
but it is actually recovered in the solution.
Regarding the range of possible values of $\kappa_o$, this parameter is bounded 
by below, $\kappa_o\geq1/8$. In particular, this lower bound is associated
with the vanishing of the lapse function on the horizon. No maximum
value exists for $\kappa_o$. However, the  quadratic growth of the
lapse when $\kappa_o$ increases, makes $N$ to reach very large values 
very rapidly. For this reason the convergence of our codes is limited
to a maximum value of  $\kappa_o$.

\begin{figure}[t]
	\centerline{\includegraphics[scale=0.7]{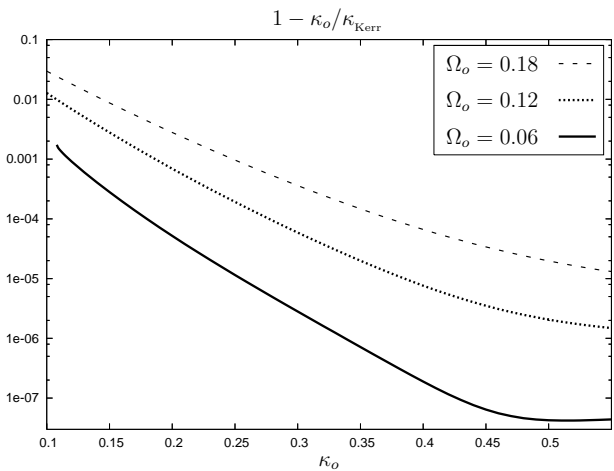}}
	\caption[]{
		\label{f:kkerr_kappao}  
		Relative difference between $\kappa_{_{\mathrm{Kerr}}}(a,J)$ and
		$\kappa_o$, as a function of $\kappa_o$,
		for solutions to ($b=N, \kappa = \kappa_o$) for different values of
		$\Omega_o$ in the rotating case.
	}
\end{figure}

\noindent More generally, we have also found that the 
system ($b=N, \kappa = \kappa_o$) is well-posed in the rotating case,
$\Omega_o\neq0$.
On the other hand, the situation
regarding the canonical value of the surface gravity changes with 
respect to the spherically symmetric case: the {\it a posteriori} {}
evaluation of $\kappa_{_{\mathrm{Kerr}}}(a,J)$ does not provide
identically $\kappa_o$. 
This means that, in case of existing a solution to the system 
$[b=N, \kappa=\kappa_{_{\mathrm{Kerr}}}(a,J)]$, the problem is 
not infinitely degenerated. In particular
such a solution only exists if, when screening the solutions obtained 
by prescribing $\kappa$, there is a value $\kappa_o$ for which 
$\kappa_{_{\mathrm{Kerr}}}(a,J)=\kappa_o$ holds. Our numerical
implementations seem to rule out 
this possibility, as illustrated  by Fig. \ref{f:kkerr_kappao}.
This figure (implemented in conformal flatness, for concreteness)
shows the relative difference between the prescribed value
of $\kappa_o$, and the evaluation of $\kappa_{_{\mathrm{Kerr}}}(a,J)$
in the constructed solution. As in the spherically symmetric case, a
minimum value for $\kappa_o$ is found, whereas the growth of the
lapse limits the upper values we can numerically implement.
In this range, and for $\Omega_o=0.06$, a non-vanishing minimum
difference between  
$\kappa_o$ and $\kappa_{_{\mathrm{Kerr}}}(a,J)$ is actually found.
A similar behaviour is found for bigger values of $\Omega_o$, even
though the numerical limitations prevent us from determining the minimum.

Therefore in the rotating case $\Omega_o\neq 0$ the system [$b=N, \kappa=\kappa_{_{\mathrm{Kerr}}}(a,J)$]
seem also to be ill-posed, but for the non-existence
of solutions rather than because of the presence of infinitely many.

\subsection{Extending the well-posedness analysis from CTT to CTS: 
  $\Psi^6\cdot K_{ij}s^is^j$ versus $\theta_{(\hat{k})}$ prescription}
One of the main goals of this work is the study of condition
(\ref{Kss_ineq}), (\ref{e:BC_btilde_mixed}) in the CTS system. In Ref. \cite{DainJK05} it was shown that
this condition, together with (\ref{e:BC_Psi}) and (\ref{e:shearzero}),
defined a well-posed problem in the CTT construction of initial data.
A natural question is to study how this result extends when the elliptic 
system is enlarged with an additional equation for the lapse. 
In a strict sense, this question is not properly formulated,
since its answer will depend on the fifth boundary condition ``for the
lapse''. The aim here is rather to assess if a qualitative conclusion
(independent of the details of the fifth boundary condition),
can be formulated about the possible range of values of 
$\Psi^6\cdot K_{ij}s^is^j$, with a focus on the negative ones.
Our intention is not
to perform an academical (numerical) extension of the CTT analytical result: 
we are ultimately motivated by probing some 
technical issues that will arise in the dynamical regime of the horizon.
Indeed, given the relation with $\theta_{(\hat{k})}$ via Eq. (\ref{e:BC:fut_marg_trapped}), and
the need to control the sign of  $\theta_{(\hat{k})}$ in a dynamical horizon,
it is fundamental to know the range of values we can actually
prescribe on ${\cal S}_t$.

\begin{figure}[b]
	\centerline{\includegraphics[scale=0.7]{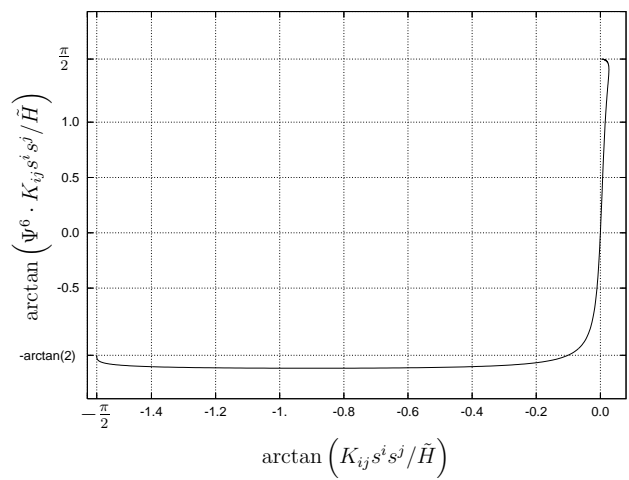}}
	\caption[]{
		\label{f:thetak-psikss} 
		Numerical analysis of condition (\ref{Kss_ineq}). 
		Values of $\theta_{(\hat{k})}/\tilde{H}= K_{ij}s^is^j/\tilde{H}$
		and $\Psi^6\cdot K_{ij}s^js^j/\tilde{H}$ in the spherically symmetric case, after
		compactification by means of the $\mathrm{arctan}$ function.
	}
\end{figure}

\medskip

\noindent {\it a) Non-rotating case: $\Omega_o=0$}. 
Let us firstly consider the spherically symmetric case.
In addition to the
boundary conditions in the first two columns of Table
\ref{t:method_BC}, 
and in order to probe the possible values of $\Psi^6\cdot
K_{ij}s^is^j$, let us 
prescribe $K_{ij}s^js^j=\lambda<0$, with $\lambda$ a negative constant
on ${\cal S}$.
We then complete the elliptic system with different boundary
conditions ``for the lapse'', and for each of them 
we screen the (negative)  values of $\lambda$ for which we can construct a solution. 
Finally for each found solution, i.e. for each value of $\lambda$,  
we plot 
the dimensionless
quantity $(\Psi^6\cdot K_{ij}s^is^j)/(\tilde{H})$ against 
$(\theta_{(\hat{k})})/(\tilde{H})$.
In  Fig. \ref{f:thetak-psikss} we present the 
resulting curve 
where, for completitude, an extension to positive values of $\lambda$
has been included. 
The curve proves to be independent 
of the ``lapse'' boundary condition and,
even though numerically we can only reach finite
values of $\lambda$, in this spherically symmetric case an
analytical expression can in fact be obtained (see Appendix \ref{s:app}). 
For this reason, a compactification with the function
$\mathrm{arctan}$
has been implemented for plotting Fig. \ref{f:thetak-psikss}.

The existence in Fig. \ref{f:thetak-psikss}
of a minimum $\delta_{min}$ for $\Psi^6\cdot
K_{ij}s^is^j$ and of an asymptotic negative non-vanishing value $\delta_{asym}<0$ 
for  $\Psi^6\cdot K_{ij}s^is^j$ (when $K_{ij}s^is^j\to
-\infty$), show that: i) the possible negative values of $\Psi^6\cdot K_{ij}s^is^j$ are bounded
by below: $\delta_{min}\leq (\Psi^6\cdot K_{ij}s^is^j)/(\tilde{H})$;  and ii) 
there is a range $\delta_{asym}\leq(\Psi^6\cdot
K_{ij}s^is^j)/(\tilde{H})\leq 0$ for which the prescription of
the value of  $\Psi^6\cdot
K_{ij}s^is^j$ determines a unique solution for the
elliptic system, exactly as it was concluded analytically in the CTT case.
Let us also note that uniqueness is lost when the value  $\Psi^6\cdot
K_{ij}s^is^j$ is prescribed between the values
$\delta_{min}<\Psi^6\cdot K_{ij}s^is^j<\delta_{asym}$, for which two solutions exist.
This is just another example of the
non-uniqueness issue associated with solutions of the CTS equations, already pointed out in Ref.
\cite{PfeifY05}. 
This is in contrast with the prescription 
of a negative non-conformally rescaled  
$K_{ij}s^is^j=\theta_{(\hat{k})}$: in the spherically symmetric case,
for each negative 
value of $\theta_{(\hat{k})}$ (not bounded by below) 
there exists a unique solution.
This suggests $\theta_{(\hat{k})}$ as
the good function to be prescribed on ${\cal S}$,
since no knowledge additional to the sign of $\theta_{(\hat{k})}$
is needed for consistency   
(see below).

The situation is reversed for positive values of $\lambda$,
where only small values for  $\theta_{(\hat{k})}$ can be prescribed,
leading always to a degenerate solution. The good parameter is then
$\Psi^6\cdot K_{ij}s^is^j$. However, these solutions do not describe
a {\it future} marginally trapped surface and their interest in the
present context is only formal.

\medskip

\begin{figure}[t]
	\centerline{\includegraphics[scale=0.7]{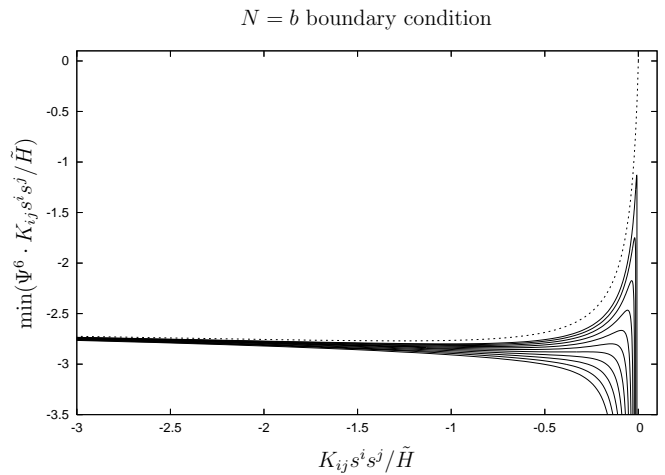}}
	\caption[]{
		\label{f:Fig2_Rot_N_eq_b} 
		Values of  $\Psi^6\cdot K_{ij}s^is^j/\tilde{H}$ and $\theta_{(\hat{k})}/\tilde{H}$,
		for boundary conditions $ K_{ij}s^is^j=\lambda$, $b=N$ 
		and $\Omega_o = 0, 0.10, 0.11, 0.12, ..., 0.20$.
		Curves depart from the spherically 
		symmetric case (dotted curve corresponds to $\Omega_o=0$) as the rotating parameter increases. 
		Solutions exist for every precribed $\theta_{(\hat{k})}$ and
		are unique. 
	}
\end{figure}

\begin{figure}[t]
	\centerline{\includegraphics[scale=0.7]{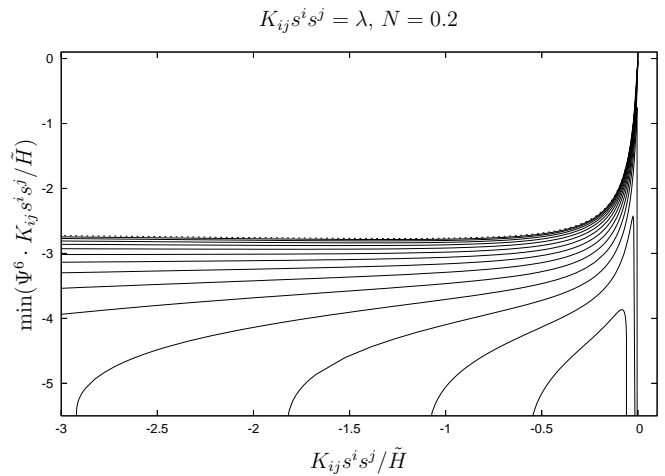}}
	\caption[]{
		\label{Fig2_Rot_N_eq_0.2} 
		Values of  $\Psi^6\cdot K_{ij}s^is^j/\tilde{H}$ and $\theta_{(\hat{k})}/\tilde{H}$,
		for boundary conditions $ K_{ij}s^is^j=\lambda$, $N=0.2$ 
		and $\Omega_o = 0.00, 0.01, 0.02, ..., 0.15$. 
		If the rotating parameter is smaller than $\Omega_o \approx 0.12620$, the
		curves reach the coordinate origin, otherwise they diverge to
		$-\infty$ as $\theta_{(\hat{k})}/\tilde{H}\rightarrow 0$.
	}
\end{figure}

\begin{figure}[t]
	\centerline{\includegraphics[scale=0.7]{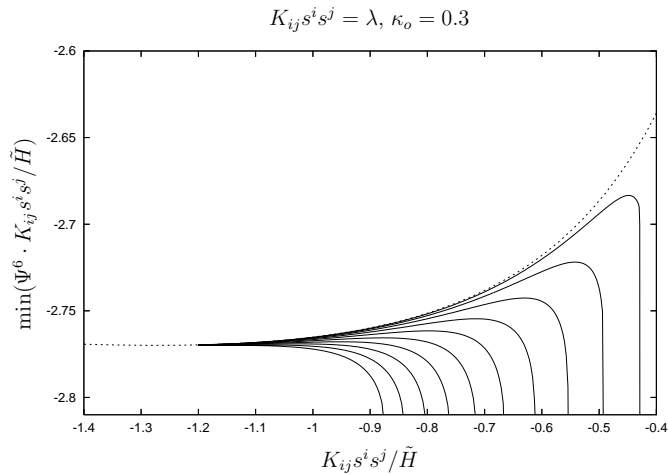}}
	\caption[]{
		\label{f:Fig2_Rot_kappa_const} 
		Values of  $\Psi^6\cdot K_{ij}s^is^j/\tilde{H}$ and $\theta_{(\hat{k})}/\tilde{H}$,
		for boundary conditions $K_{ij}s^is^j=\lambda$, $\kappa_o=0.3$
		and $\Omega_o = 0.00, 0.02, 0.04,
		..., 0.18, 0.20$.  
		Curves
		with $\Omega_o\neq 0$ do not pass through the origin:
		conclusions in Ref. \cite{DainJK05} do not extend to the rotating case. On
		the other hand, $\Omega_0\neq 0$ curves become indistinguishable of the spherically
		symmetric curve (dotted line) for
		sufficiently negative values of $K_{ij}s^is^j$.
	}
\end{figure}

\noindent {\it b) Rotating case: $\Omega_o\neq0$}. 
The independence of the curve 
$\Psi^6\cdot K_{ij}s^is^j$ vs. $\theta_{(\hat{k})}$ with respect to 
the ``lapse boundary condition'' disappears in the rotating case.
This is illustrated in Figs. \ref{f:Fig2_Rot_N_eq_b},
\ref{Fig2_Rot_N_eq_0.2}  and \ref{f:Fig2_Rot_kappa_const} , 
where different boundary conditions for $N$ have been
implemented\footnote{\label{umbilical}In order to implement quasi-equilibrium, and according
to  Eq. (\ref{e:shearzero}),
we restrain here to a flat
$\tilde{\gamma}_{ij}$ (this guarantees the umbilical condition 
$\tilde{H}^{ij} -\frac{1}{2} \tilde{q}^{ij}
\tilde{H}=0$).} (due to the
angular dependence, we plot now the minimum of $\Psi^6\cdot
K_{ij}s^is^j$).

The most dramatic qualitative change in the rotating case is the
existence of a certain value of
$\Omega_o$ such that, for bigger values of this rotation parameter, 
the curve does not pass through the
origin (in the case of Figs. \ref{f:Fig2_Rot_N_eq_b} and  \ref{f:Fig2_Rot_kappa_const} this is true for any $\Omega_o\neq
0$). Arbitrarily negative values of 
$\Psi^6\cdot K_{ij}s^is^j$  are found for small values of
$\theta_{(\hat{k})}$. Therefore in this range of $\Omega_o$, and in contrast
with the spherically symmetric case, the negative values of 
$\Psi^6\cdot K_{ij}s^is^j$ are only bounded from
{\it above}. This shows that conclusions in Ref. \cite{DainJK05} for
the CTT case {\it do not extend} to the CTS one for arbitrary values
of $V^i = \Omega_0 \cdot \phi^i$.
Given the variety of behaviours in
Figs. \ref{f:Fig2_Rot_N_eq_b}, \ref{Fig2_Rot_N_eq_0.2} and \ref{f:Fig2_Rot_kappa_const},
it is difficult to extract generic conclusions,
i.e. independent of the boundary conditions on $N$, about the
possible values of $\theta_{(\hat{k})}$ and $\Psi^6\cdot K_{ij}s^is^j$.
Our numerical simulations do not provide a
complete understanding, but rather some restricted insight, of the
boundary conditions properties in the case 
of a rotating black hole.

Having stated this clearly, we highlight Fig. \ref{f:Fig2_Rot_N_eq_b}
corresponding to the interpretation of boundary condition (\ref{b=N})
as the ``lapse'' boundary condition, $N=b$.
In this case, 
we can actually formulate a concrete statement  about 
the good parameter to be prescribed:
for all values of $\Omega_o$, there exists a (small) negative value $K_{ij}s^is^j=
\Delta_{\Omega_o}\leq 0$ 
such that the prescription of  
$\theta_{(\hat{k})}$ to $\lambda<\Delta_{\Omega_o}$ determines a unique solution.
It is very difficult to determine numerically if $\Delta_{\Omega_o}/\tilde{H}$ is
actually zero or a very small value depending on $\Omega_o$. At this value $K_{ij}s^is^j=\Delta_{\Omega_o}$
the conformal factor seems to have a pole and the quantity
$\Psi^6\cdot K_{ij}s^i s^j$
diverges to negative values. In consequence curves in Fig.
 \ref{f:Fig2_Rot_N_eq_b} for $\Omega_o\neq 0$ do not reach the origin,
as pointed out above, changing the qualitative behaviour discussed in the spherically 
symmetric case. Still there exists a critical
$\Omega_{crit} (\approx 0.15)$ such that, in the range $\Omega_o<\Omega_{crit}$,
all curves present a local minimum $\delta_{min}$ and an asymptotic value
$\delta_{asym}$ at $K_{ij}s^is^j\to -\infty$. 
However,  uniqueness (if existence at all) is lost 
when prescribing $\Psi^6\cdot K_{ij}s^i s^j > \delta_{asym}$: 
because of the negative divergence of $\Psi^6\cdot K_{ij}s^is^j$, either there
are two or none solutions.
The ultimate reason for focusing on this ``fifth'' boundary condition,
is that it presents some geometrical/physical
advantage with respect to the other ones: i) it implements the NEH
condition for any choice of $\tilde{\gamma}_{ij}$ without the need of relying on
the umbilical condition (see footnote
\ref{umbilical})
and ii) by enforcing the evolution vector $t^\alpha$ to be tangent to
$\Hor$, which is in quasi-equilibrium, it chooses a coordinate system 
on the horizon in which no time-dependence 
is artificially introduced as a gauge effect
(see also footnote
\ref{b=Nquasi-eq} for actual physical consequences in the binary case).

In sum, we conclude that $\Psi^6\cdot K_{ij}s^is^j$ is not
a good function to be prescribed in a CTS approach. 
The present numerical analysis suggests that adaptation of the
coordinate system to the horizon (boundary condition $b=N$) and
the prescription of $\theta_{(\hat{k})}$ to a sufficiently negative value
determines a unique solution.
This is a relevant information
for the dynamical\footnote{In the dynamical case, the boundary condition
$b=N$ is generalized by solving an elliptic equation on ${\cal S}$; see
\cite{Eardl98,GourgJ06c}.} case, in particular
in a constrained evolution scheme in which the elliptic
system (\ref{CTSeqs}) is solved at each time step
(see Ref. \cite{BonazGGN04}).

\subsection{Prescription of the divergence of $\Omega_i$}
Since the solution $\mathrm{ln}N_0$ of Eq. (\ref{Domega})
is defined up to a constant on ${\cal S}$, the value 
$N=C\cdot e^{\mathrm{ln}N_0}$ provides a Dirichlet condition
for the lapse, which takes into account the manner in which ${\cal S}$
is embedded in the spacetime and permits to control the magnitude of $N$
via the free choice of the constant $C$.
The multiplicative character of the latter is a good feature on
numerical grounds. 
Of course, condition (\ref{Domega})
must be completed with a choice of ${}^2D^k\Omega_k$ (an
arbitrary function on ${\cal S}\approx S^2$ with vanishing $\ell=0$ mode). This makes 
this condition an {\it effective} one, in the sense discussed in Ref.
\cite{CookP04} for ``lapse boundary conditions''. 
However, for specific problems natural choices exist, e.g. ${}^2D^k\Omega_k=0$
for perturbations of the spherically symmetric case (see \cite{AshteK05,GourgJ06} 
for a {\it Pawlowski gauge}, motivated by Kerr-Schild-like slicings).
More importantly, a part of the extrinsic geometry is incorporated into this
boundary condition, which thus goes beyond a merely numerically convenient
choice. Once ${}^2D^k\Omega_k(=0)$
has been chosen, the slicing is completely fixed, and
all the freedom is reduced to a single constant $C$
that controls the rate to move across the slicing.
We have numerically verified the existence of unique solutions when combining this condition
with those ones in the third column of Table \ref{t:method_BC}.

\subsection{General Comments}
\noindent{\it Other possible combinations}.
Table \ref{t:method_BC} presents an association between 
boundary conditions and constrained fields. Even though 
this can be useful to organize the discussion, it must again be
emphasized that this is 
only a methodological option. 
Insisting on such an association can be misleading
and can shadow some useful choices. As an example, 
since condition (\ref{e:BC_Psi}) provides
an expression for $K_{ij}s^is^j$, it 
could be interpreted as a condition\footnote{In particular, 
together with (\ref{e:shearzero}) as a condition on $V^i$,
this means that non-expanding conditions can be fully
fulfilled by an appropriate choice of the shift.} 
for $\tilde{b}$
via Eq. (\ref{e:BC_btilde_mixed}).
Then, interpreting $b-N=0$ as a condition for $N$, we can think of
prescribing a value for $\Psi$ on ${\cal S}$ by a Dirichlet 
condition. This particular example has two 
relevant applications:
{\it i)} it permits to prescribe the area of ${\cal S}$ 
(since the conformal part is a free data) and {\it ii)} it provides
an alternative strategy for the analytical study of the well-posedness
of the marginally trapped condition (Refs. \cite{Maxwe04,Dain04} focus
on the control of the radial derivative of $\Psi$ in
order to apply a maximum principle to the Hamiltonian constraint equation; the
 alternative of a Dirichlet condition for $\Psi$ 
shifts the focus to the momentum constraint).

\noindent{\it Effective boundary conditions}. At the end of the day, 
quasi-equilibrium conditions (\ref{NEHgeom}) together with 
$b-N=0$ leave a single function to be specified 
on ${\cal S}$. Following  \cite{Cook02,JaramGM04,CookP04}
this fifth function can be seen as a condition ``for the lapse'',
either geometrically motivated or purely {\it effective}.
But it must be underlined (see also in this sense the 
discussion in section III.C of Ref. \cite{CookP04}),
that it can be more generally interpreted 
as an {\it effective} condition on any scalar
combination of fields, e.g. the conformal factor or
${}^2D^k\Omega_k$.

\noindent{\it Generic well-posedness}. Finally, we have commented that combining
conditions in Table \ref{t:method_BC} lead to well-posed problems
for generic choices of $f_1$ and $f_2$. By this we mean that only for some
critical  $f_1$ and $f_2$ the problem admits an infinite number of solutions.
For instance, condition ${\cal L}_{\ell}\theta_{(k)}=0$ in
\cite{Cook02} can be recasted
as a prescription $f_1$ for $\Psi^6 \cdot K_{ij}s^is^j$, and also as 
${}^2D^k\Omega_k=f_2$. If these particular expressions are used 
for conditions  (\ref{e:BC_btilde_mixed}) or (\ref{Domega}) in the
spherically symmetric case, this leads to an ill-posed problem,
as shown in Ref. \cite{CookP04}.
By {\it generic} me mean here that a small perturbation of this critical cases
makes the problem well-posed.

\section {Conclusions}
This work represents the numerical counterpart of Refs. \cite{JaramGM04,DainJK05,GourgJ06},
where isolated horizon boundary conditions were proposed. As a first result,
the prescription of a constant value $\kappa=\kappa_o$ on a instantaneously  
non-expanding horizon, using a coordinate system adapted to the horizon
[i.e. conditions (\ref{kappa=const}), (\ref{NEHgeom}) and (\ref{b=N})], defines a well-posed problem.
If $\kappa$ is set to $\kappa_{_{\mathrm{Kerr}}}(a,J)$, as proposed in  \cite{JaramGM04},
the problem is degenerated (infinite number of solutions)
in the spherically symmetric case and admits no solution when rotation is
introduced. The only freedom in this system is the choice of the constant $\kappa_o$.
The $\kappa=\mathrm{const}$ condition does not enforce a quasi-equilibrium
restriction on the geometry of the horizon. This means that it does not increase the
physical degree of stationarity of the associated
initial data (consistently with \cite{CookP04}, where it is shown that
physical quantities do not depend on the chosen {\it lapse} boundary condition). 
Its interest is rather in the evolution of such initial data, since it provides a slicing where the lapse 
function must be initially time independent, something desirable numerically.

Secondly, the results in Ref. \cite{DainJK05} on the prescription of $\Psi^6\cdot
K_{ij}s^is^j$ in the CTT system, do not extend
straightforwardly to the CTS case, except in spherical symmetry.
In this particular case,  there exists a negative bound $\delta_{asym}$ 
such that for $\delta_{asym}\leq\Psi^6\cdot K_{ij}s^is^j/\tilde{H}\leq0$ 
there is a unique solution. There exists a second bound $\delta_{min}$ such
that the prescription
$\delta_{min}<\Psi^6\cdot K_{ij}s^is^j/\tilde{H}<\delta_{asym}$ 
admits two degenerate solutions.
For $\Psi^6\cdot K_{ij}s^is^j/\tilde{H}<\delta_{min}$ no solution exists.
In the rotating case, the strong dependence on the fifth boundary condition
prevents us from deriving general bounds for
$\Psi^6\cdot K_{ij}s^is^j$. 
However, for the particular choice of a coordinate system 
adapted to the horizon ($b=N$ condition), the prescription 
of $\theta_{(\hat{k})}$ to a sufficiently negative value guarantees
the existence and uniqueness of a solution to the CTS elliptic system.
Therefore, rather than $\Psi^6\cdot K_{ij}s^is^j$, the good parameter to 
be prescribed is $\theta_{(\hat{k})}$. 
This represents an important information for 
the implementation of evolving black holes as 
{\it regular future} trapping/dynamical horizons \cite{AshteK05,Booth05,Simon06}
in a constrained evolution scheme.

Finally we have underlined the fact that prescribing 
non-expanding and adapted-coordinate-system conditions leave one
free function to be specified on the horizon. Due to the (non-linear) coupled nature of
boundary conditions, this fifth condition is not specifically related to 
a particular field and, even though it can be useful 
to interpret it as a lapse boundary condition, other choices 
(e.g. a Dirichlet condition on $\Psi$) can prove to be useful.

\section{Acknowledgements}
The authors are grateful to S. Bonazzola, S. Dain, E. Gourgoulhon, B. Krishnan,
G. Mena-Marug\'an and J. Novak for numerous discussions.
JLJ acknowledges the support of the Marie
Curie Intra-European contract MEIF-CT-2003-500885 
within the 6th European Community Framework Programme, and the
hospitality of the Albert Einstein Institute. 

\appendix

\section{Comments on Figure \ref{f:thetak-psikss}}
\label{s:app}
In this appendix we derive some analytical relations that are valid
for CTS data in the conformally flat, maximal, spherically symmetric
case. In particular we prove that
the graph displayed in Fig. \ref{f:thetak-psikss} 
is independent of a specific choice of inner boundary conditions on 
$N$ and $\tilde{b}$ (the prescription $K_{ij}s^js^j=\lambda$ determine
one point in the curve), and give a parametric 
analytical expression through which this curve is defined.

\paragraph{Independency of Fig.\ref{f:thetak-psikss}}.
In spherical symmetry, with $(r,\theta,\phi)$ being spherical coordinates in 
which the apparent horizon is located at $r=r_H$, the  maximal slicing
CTS equations 
(\ref{CTSeqs}) [with $a=0$] reduce to:
\bea
&&	\label{psi_rr}
	\frac{d^2\Psi}{dr^2} + \frac{2}{r}\frac{d\Psi}{dr}
		+ \frac{\Psi^5}{12N^2} r^2 \left[\frac{d}{dr}\left(r^{-1}\beta^r\right)\right]^2 = 0 \\
&&	\label{beta_rr}
	\frac{d^2}{dr^2}\left(r^{-1}\beta^r\right) + 
		\frac{d}{dr}\left(r^{-1}\beta^r\right)\left[\frac{4}{r}-\frac{d}{dr}\left[\log\left(N\Psi^{-6}\right)\right]\right] 
		= 0 \nn \\
&&	\label{alpha_rr}
	\frac{d^2}{dr^2}(N\Psi) + \frac{2}{r}\frac{d(N\Psi)}{dr}
		-
		\frac{7\Psi^5}{12 N}r^2\left[\frac{d}{dr}\left(r^{-1}\beta^r\right)\right]^2  = 0 \nn
\eea
and the apparent horizon boundary condition (note that
$\beta^\theta=\beta^\phi=0$) is given by
\beq
	\label{AppHor_r}
	\left[\frac{d\Psi}{dr} + \frac{\Psi}{2r} + \frac{r\Psi^3}{6N}
	  \frac{d}{dr}\left(r^{-1}\beta^r\right)\right]_{r=r_H} = 0 \ \ .
\eeq
With the introduction of the compactified radial coordinate $s=r_H/r$,
we get
\bea
	\label{psi_ss1}
&&	s^2\Psi''       +      B'^2\frac{\Psi^5}{12 N^2}        = 0 \\
	\label{beta_ss}
&&	B''             -      B'\left(\frac{2}{s}+\log\left[N\Psi^{-6}\right]'\right) = 0 \\
	\label{alpha_ss}
&&	s^2(N\Psi)''    -    7B'^2\frac{\Psi^5}{12 N^2}           = 0 
\eea
and
\beq
	\label{AppHor_s1}
	\left[\Psi' - \frac{\Psi}{2} + \frac{\Psi^3}{6N}B'\right]_{s=1} = 0\,,
\eeq
where $' = d/ds$ and $\beta^r= B/s$.
Equation (\ref{beta_ss}) can be solved explicitly:
\beq
	\label{Sol_dB}
	B'= c \cdot s^2 N \Psi^{-6}\,,
\eeq
where $c$ is a constant of integration, closely related to
$\Psi^6\cdot K_{ij}s^is^j$ [see Eq. (\ref{e:BC_btilde_mixed}) and below]. Using (\ref{Sol_dB}) we obtain 
\beq
	\label{psi_ss2}
	\Psi''       +      \frac{c^2}{12}s^2\Psi^{-7}   =0
\eeq
together with the boundary condition 
\beq
	\label{AppHor_s2}
	\left[\Psi' - \frac{\Psi}{2} + \frac{c}{6\Psi^3}\right]_{s=1} = 0\,.
\eeq
The system (\ref{psi_ss2}, \ref{AppHor_s2}) uniquely defines a
sequence of solutions $\Psi(s;c)$
which is independent of a specific choice of inner boundary conditions
for $B$ and $N$.
Moreover, the horizon quantity (cf. Eq. (\ref{e:BC_btilde_mixed}) in
the spherically symmetric case)
\bea\label{Kss_1}
	&&\left[K_{ij}s^i s^j \right]_{r=r_H} \\
&& = \left[\frac{2}{3N}\left(\frac{d}{dr}\beta^r-\frac{1}{r}\beta^r\right)\right]_{r=r_H}
		=-\frac{2}{3r_H}c\Psi_H^{-6}\,, \nn
\eea
where
\beq\label{psiH_of_c}
	\Psi_H = \Psi(s=1;c) \ \ ,
\eeq
is a function of $c$ alone. Thus also the graph displayed in Fig. $\!{}$\ref{f:thetak-psikss}
is independent of a specific choice of inner boundary conditions on $\beta^r$ and $N$.
The prescription $K_{ij}s^is^j=\lambda$ determines a point in this curve.

\paragraph{Analytic representation of  Fig. $\!{}$\ref{f:thetak-psikss}.}
In order to obtain explicit mathematical expressions that describe the graph displayed in  Fig. $\!{}$\ref{f:thetak-psikss}, 
we consider the spatial metric as well as the extrinsic curvature of the family of 
time-independent maximal slicings of the Schwarzschild solution \cite{EstabW73}, i.e.
\beq\label{Esta_ds2_1}
	ds^2 = \left(1-\frac{R_H}{R}+C^2\frac{R_H^4}{R^4}\right)^{-1} dR^2 + R^2(d\theta^2+\sin^2\theta d\phi^2)
\eeq
and
\beq\label{Esta_K}
	K^i_j=C\;\frac{R_H^2}{R^3}\left(
	\begin{array} {ccc} 
		-2 & 0 & 0 \\
		 0 & 1 & 0 \\
		 0 & 0 & 1 
	\end{array}\right)\,,
\eeq
which gives in particular
\beq\label{Kss_2}
	\left[K_{ij}s^i s^j \right]_{R=R_H}= -\frac{2C}{R_H}\,.
\eeq
In these coordinates $(R,\theta,\phi)$, the radius of the black hole horizon is given by $R=R_H=2M$, where $M$ is the black hole mass. The constant $C$ parametrizes the family of maximal slicings.

The coordinate transformation $r=r(R)$,
leading to a conformally flat line element
\beq\label{Esta_ds2_2}
	ds^2 = \Psi^4\left[dr^2 + r^2(d\theta^2+\sin^2\theta
	  d\phi^2)\right] \ \ ,
\eeq
is described by
\beq\label{s_of_S}
	s = \exp\left(\int_1^S \frac{d\sigma}{\sigma\sqrt{1-\sigma+C^2\sigma^4}}\right)\,,
\eeq
where
\beq
	s = \frac{r_H}{r} \qquad \mbox{and} \qquad S = \frac{R_H}{R} \,.
\eeq
The conformal factor $\Psi$ which satisfies the boundary value problem (\ref{psi_ss2}, \ref{AppHor_s2}),
can be obtained from the comparison of the line elements (\ref{Esta_ds2_1}, \ref{Esta_ds2_2}):
\beq
	\Psi^2 = \frac{R}{r}=\frac{R_H}{r_H}\frac{s}{S}\,.
\eeq
As a consequence
\beq
	\Psi_H^2 = \frac{R_H}{r_H}\,,
\eeq
and by imposing the asymptotic boundary condition, $\lim_{S\to 0} \Psi = 1$,
we can write $\Psi_H^2$ in terms of the parameter $C$:
\beq\label{psiH_of_C_pos}
	\Psi_H^2(C\geq 0)
	= \exp\left(\int_0^1\frac{d\sigma}{\sigma}\left[\frac{1}{\sqrt{1-\sigma+C^2\sigma^4}}-1\right]\right)\,.
\eeq
The formula (\ref{psiH_of_C_pos}) only describes $\Psi_H$ for
non-negative values of $C$. An expansion of this 
expression into the realm of negative $C$-values yields, for $0>C>-3\sqrt{3}/16$
\bea\nonumber
\Psi_H^2(C<0)&=&
	\label{psiH_of_C_neg}
	\exp\left(\int_0^1\frac{d\sigma}{\sigma}\left[\frac{1}{\sqrt{1-\sigma+C^2\sigma^4}}-1\right] \right.\\
&&\left. + 2\int_1^{s_1(C)}\frac{d\sigma}{\sigma
	  \sqrt{1-\sigma+C^2\sigma^4}}\right)\label{psiH_of_C_neg} \ ,
\eea
where $s_1(C)$ is the real zero of 
\[
	f(s) = 1-s+C^2 s^4 \ \ ,
\]
with 
\[
	1\leq s_1(C) \leq \frac{4}{3}\,.
\]
Note that $\Psi_H^2(C)$ tends to $+\infty$ as $C \to -3\sqrt{3}/16$ since $f(s)$ has a double zero, $s_1=4/3$, in this limit.

With (\ref{Kss_2}) and $\tilde{H}=2/r_H$, we 
finally obtain the desired parametric description of the curve
displayed in  Fig. $\!{}$\ref{f:thetak-psikss}
[a parametrization in terms of $c$ in Eq. (\ref{psiH_of_c}) can be also
obtained from the relation between $C$ and $c$ provided by
Eqs. (\ref{Kss_1}) and (\ref{Kss_2})]:
\beq
\begin{array}{cccc}
	\mbox{Abscissa:}&\quad K_{ij}s^i s^j/\tilde{H} &=& -C\Psi_H^{-2} \\[5mm]
	\mbox{Ordinate:}&\quad \Psi_H^6\cdot K_{ij}s^is^j/\tilde{H} &=& -C\Psi_H^{4}
\end{array}
\eeq
A particular consequence of this analysis is 
\beq
	\lim_{C\to\infty} \Psi_H^6\cdot K_{ij}s^i s^j = -2\tilde{H}\,,
\eeq
providing the asymptotic value when $K_{ij}s^i s^j/\tilde{H} \to -\infty$.

\end{document}